\documentclass[aps,prb,onecolumn,,floatfix]{revtex4-2}

\usepackage{amsmath}
\usepackage{amssymb}
\usepackage{graphicx, float}

\newcommand{\bea}{\begin{eqnarray}}
\newcommand{\eea}{\end{eqnarray}}

\newcommand{\parent}[1]{\left( #1 \right)}

\begin{document}


\title{Minimal entropy production under thermodynamic constraints: \\
An application of cycle decomposition for Markov chain design}

\author{David Andrieux}

\noaffiliation


\begin{abstract}
We propose to construct Markov dynamics with specified characteristics using a cycle decomposition of the space of Markov chains introduced by Cohen \cite{C81} and Alpern \cite{A83}.
As an application of this approach, we derive the minimal entropy production required to generate prescribed thermodynamic currents.
\end{abstract}


\maketitle

\section{Context and objectives}

Experimental advances open the way for refined control of mesoscopic systems, from molecular motors to electronic quantum junctions.
The next frontrier will be to design these systems so that they exhibit specific dynamical or thermodynamical properties.
Examples of systems where dynamical properties can increasingly be controled include the growth and composition of multi-component structures \cite{WSH12, GA14, AG13}. 

However, designing dynamics with given properties remains challenging. 
Part of the issue comes from the limited available analytical results.
In addition, a more structural challenge comes from the difficulty to systematically explore the space of stochastic matrices. 

In this paper we propose that a decomposition of Markov dynamics introduced by Cohen \cite{C81} and Alpern \cite{A83} can be used to further study, and ultimately design, Markov dynamics. 
Cohen and Alpern showed that Markov dynamics can be expressed as a convex combination of cycle matrices. 
Using this approach, it is possible to construct dynamics with given properties, such as a given steady state distribution. 
Another application is computing the nearest reversible Markov chain \cite{NW12}. 
A possible extansion of this approach would be to construct the set of dynamics with fully symmetric nonlinear response \cite{A22, A23b}. 

Here we show how this approach can help understand and design systems by investigating the entropy production under constraints. 
Specifically, we derive the minimal entropy production required to sustain thermodynamic currents in an arbitrary topology of the transition network.


\section{Decomposition of the space of Markov chains in cycle matrices}

We consider a Markov chain characterized by a transition matrix $P = \parent{P_{ij}} \in \mathbb{R}^{N\times N}$ on a finite state space.
We assume that the Markov chain is primitive, i.e., there exists an $n_0$ such that $P^{n_0}$ has all positive entries. 
The chain $P$ thus admits a unique stationary distribution $\pi$.

It will be convenient to refer to the associated matrix $F$ such that $F_{ij} = \pi_i P_{ij}$. 
Each element $F_{ij}$ corresponds to the steady state probability flux between state $i$ and $j$.
$F$ satisfies 
\bea
\sum_i F_{ij} = \pi_j \quad \text{and} \quad \sum_j F_{ij} = \pi_i
\label{ss}
\eea
such that $\sum_{ij} F_{ij} =1$. Every matrix $F$ corresponds to a unique chain $P$ and vice versa. 

Cohen \cite{C81} and Alpern \cite{A83} demonstrated that the matrix $F$ can always be decomposed as a combination of {\it cycle matrices}. 
If $[a_1, \cdots, a_m]$ is a sequence of distinct integers chosen from $1, \cdots, N$, then we define the corresponding cycle matrix as the $N \times N$ matrix $C^{[a_1, \cdots, a_m]}$ given by $c_{a_1a_2} = c_{a_2a_3} = \cdots = c_{a_m a_1} = 1/m$ and $0$ otherwise.  
We say that $\ell = m$ is the length of $C$.\\

{\bf Theorem}: Every matrix $F$ is a convex combination of cycle matrices. Furthermore, for some $n \leq N^2 - N +1$, some probability vector $(\lambda_1, \cdots, \lambda_n)$ such that $\sum_k \lambda_k =1$ and $\lambda_k \geq 0$, and some cycle matrices $C^k$, we have 
\bea
F = \sum_k \lambda_k \, C^k \, .
\label{decomp}
\eea
\\
{\it Demonstration}: See Ref. \cite{A83}.\\


Note that the decomposition (\ref{decomp}) is not unique. We will see in the next section that a thermodynamic-based decomposition will allow us to gain further insights into the behavior of Markov dynamics. 

As an illustration of Alpern's theorem, we decompose the matrix $F$ corresponding to a homogeneous random walk with current $J = (2\alpha-1)/3$ ($1/2 \leq \alpha \leq 1$):
\bea
F&=&
\frac{1}{3}
  \begin{pmatrix}
    0 &  \alpha & 1-\alpha\\
    1-\alpha & 0 & \alpha \\
    \alpha & 1-\alpha & 0
  \end{pmatrix} 
= \lambda_{[1,2,3]} C^{[1,2,3]} + \lambda_{[1,2]} C^{[1,2]} + \lambda_{[1,3]}  C^{[1,3]} + \lambda_{[2,3]} C^{[2,3]} \nonumber \\
&=& 
(2\alpha -1)
  \begin{pmatrix}
    0 & 1/3 & 0\\
    0 & 0 & 1/3 \\
    1/3 & 0 & 0
  \end{pmatrix}
+ \frac{2}{3} (1-\alpha)
  \begin{pmatrix}
    0 & 1/2 & 0\\
    1/2 & 0 & 0 \\
    0 & 0 & 0
  \end{pmatrix}
+ \frac{2}{3} (1-\alpha)
  \begin{pmatrix}
    0 & 0 & 1/2\\
    0 & 0 & 0 \\
    1/2 & 0 & 0
  \end{pmatrix}
+ \frac{2}{3} (1-\alpha)
  \begin{pmatrix}
    0 & 0 & 0\\
    0 & 0 & 1/2 \\
    0 & 1/2 & 0
  \end{pmatrix}
\nonumber
\eea
The coefficients satisfy $\lambda_k \geq 0$ and $\sum_k \lambda_k =1$. 
Notably, the coefficient $\lambda_{[1,2,3]}$ is proportional to the current $J = (2\alpha -1)/3$. We will return to this observation in the next section.

A similar decomposition applies for $0 \leq \alpha \leq 1/2$, with $\lambda_{[1,2,3]}$ replaced by $\lambda_{[1,3,2]}=1-2\alpha$ and $\lambda_{[ij]} = 2\alpha/3$. \\

The decomposition (\ref{decomp}) provides a way to construct Markov chains with specific prooperties. 
For example, we can generate all Markov chains with a specified steady state distribution by combining (\ref{decomp}) with equation (\ref{ss}) and solving the resulting linear system of equations.

The decomposition (\ref{decomp}) is also at the basis of the so-called "rotational representations" of Markov chain.
In such representations, a Markov chain is described as a partition of the unit interval, and a path trajectory of the chain is generated by the application $x_n \rightarrow (x_{n-1} +a) \mod 1$.
A rotational representation thus offer a deterministic way to simulate a Markov chain, with all the randomness captured in the initial condition \cite{G99}. 
This fact can be used to simulate trajectories without the need to generate random numbers (although we are not aware of such applications in the literature).


\section{Minimal entropy production as a function of thermodynamic currents and network topology}

We investigate the design of Markov chains that satisfy certain properties. 
In this paper we focus on Markov chains satisfying a given set of thermodynamic currents $J_\alpha$ within a given transition network.
We start by looking a dynamics with a single independent current and derive the entropy production required to generate a given current. 
We then turn to a dynamics in which two independent thermodynamic currents exist.
We finally derive the minimal entropy for an arbitrary transition network.


\subsection{Periodic ring network}

We start by looking at a Markov chain $P$ representing a ring of length $\ell = N$ with periodic boundary condition: $P_{ij} > 0$ if $|i-j|=1$ or if $|i-j|=\ell-1$, and $0$ otherwise. 
Periodic chains model many nonequilibrium systems, from the conductivity of anisotropic organic conductors to molecular motors or enzymetic kinetics. 
More generally, cycles constitute the building blocks of more complex systems, both from a dynamical and thermodynamical perspective. 
The characterization of their dynamics is thus fundamental to understand transport at the mesoscopic scale.


We will directly work with the corresponding matrix $F$. 
The steady state thermodynamic current $J = \pi_i P_{i, i+1} - \pi_{i+1} P_{i+1,1} = F_{i,i+1} - F_{i+1,1}$ 
and the affinity $A = \sum_i \ln (P_{i,i+1}/P_{i+1,i}) = \sum_i \ln (F_{i,i+1}/F_{i+1,i})$. 
In this section we will use the notation $i_{N+1} = i_1$ and $i_{0} = i_\ell$ and assume $J\geq 0$. 
The case $J \leq 0$ can be treated similarly.

We now use a result of Kalpazidou \cite{K94, K06} showing that the decomposition (\ref{decomp}) can be done using the so-called fundamental cycles of the Markov graph (see also \cite{S76} for an introduction).
As a result, we can decompose $F$ as a combination of the cycle $\alpha = [1,2,\cdots, \ell]$ and cycles $k = [k, k+1]$: 
\bea
F = \lambda_\alpha C^{\alpha} + \sum_k \lambda_k C^k 
\label{constraint1D}
\eea
where $\lambda_\alpha + \sum_k \lambda_k = 1$ (if $J$ is negative the same decomposition applies, now with $\alpha = [\ell,\cdots, 2,1]$).
In this representation, the current is given by $J = \lambda_\alpha/\ell$. 
We can thus examine all Markov chains with a given current $J$ by simply fixing the coefficient $\lambda_\alpha$. 
When $\lambda_\alpha = 0$ the matrix $F$ is symmetric and the system is at equilibrium.

We can now assess the minimal affinity (and thus entropy production) required to generate a given current $J$. 
To this end, we note that the affinity can be written as
\bea
A = \sum_k \ln \parent{\frac{F_{k, k+1} }{F_{k+1, k}}  } = \sum_k  \ln \parent{ \frac{\lambda_k/2+\lambda_\alpha/\ell }{\lambda_k/2}  } =  \sum_k \ln \parent{1+\frac{2 J}{\lambda_k}  } \,.
\label{A1}
\eea
To find the minimal affinity, we form the Lagrangian function $L$ with the Lagrange multiplier $\xi$ to impose the constraint that all $\lambda$s sum to one:
\bea
\mathcal{L} =  \sum_k \ln \parent{1+\frac{2 J}{\lambda_k}  }  + \xi \parent{\sum_k \lambda_k + \lambda_\alpha -1} \, ,
\eea
where $J = \lambda_\alpha/\ell$ is fixed. 
We now equal the derivatives with respect to $\lambda_k$ and $\xi$ to $0$ and solve the resulting system of equations.
We find $\lambda_k = (1-\lambda_\alpha)/\ell$ for all $k$, i.e. the minimum is achieved by a uniform chain \cite{FN02}.  
This leads to
\bea
A_{\min} = \ell \ln \parent{ 1 + \frac{2\lambda_\alpha}{1-\lambda_\alpha} } = \ell \ln \parent{ 1 + \frac{2\ell J}{1-\ell J} } \, .
\eea
$A_{\min}$ vanishes at equilibrium, $J = \lambda_\alpha = 0$, and diverges as the current approaches its maximal value $J = 1/\ell$ or $\lambda_\alpha =1$, as expected (as $\lambda_\alpha$ approaches unity the Markov chain becomes fully irreversible since $\lambda_k \rightarrow 0$, resulting in $P_{k,k+1}>0$ while $P_{k+1, k}=0$).

The corresponding minimal entropy production reads 
\bea
\Delta_i S_{\min} = \lambda_\alpha\times \ln \parent{ 1 + \frac{2\lambda_\alpha}{1-\lambda_\alpha} } = J \ell \times \ln \parent{ 1 + \frac{2 \ell J}{1-\ell J} } \, .
\eea
By construction this minimum can be achieved, in this case for uniform transition probabilities. 
Notably, the minimal entropy production is expressed entirely in terms of the thermodynamic current $J$ and the system size $\ell$, and is valid for any Markov chain with a ring topology.


\subsection{Network with two independent currents}

\begin{figure}[H]
\begin{center}
\includegraphics[scale=.55]{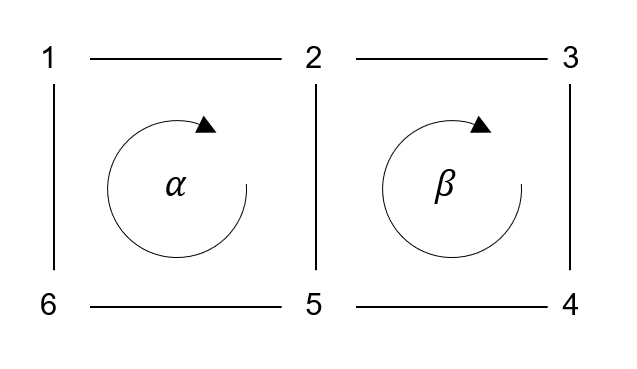}%
\caption{Transition network with two independent cycles $\alpha$ and $\beta$. The positive orientation is chosen clockwise.}
\label{Ion}
\end{center}
\end{figure}

We now consider a Markov chain with two independent thermodynamic currents $J_\alpha$ and $J_\beta$ (Figure \ref{Ion}). 
This network topology can serve, for example, as a model system of ion transport through a membrane \cite{H05}. 

For expository purposes we will look at the cases where both currents have the same or opposite directions separately. 
We will present a unified approach in the next section when analyzing the case of an arbitrary network topology. 

\paragraph{The two currents have opposite orientations.}

The currents are coupled through the transition $r = [2,5]$, and the affinities take the form
\bea
A_\alpha &=& \ln \parent{\frac{\lambda_r /2 + J_\alpha + J_\beta}{\lambda_r /2}  } + \sum_k \ln \parent{1+\frac{2 J_\alpha}{\lambda_k}  } \, , \quad  k = [5,6], [6,1], [1,2]\\
A_\beta &=& \ln \parent{\frac{\lambda_r /2 + J_\alpha + J_\beta}{\lambda_r /2 }} + \sum_l \ln \parent{1+\frac{2 J_\beta}{\lambda_l }  } \, , \quad  l = [2,3], [3,4], [4,5] \, .
\eea

We consider the minimal entropy production $\Delta_i S = J_\alpha \times A_\alpha +  J_\beta \times A_\beta$ required to produce the currents $J_\alpha$ and $J_\beta$.
We introduce the Lagrangian 
\bea
\mathcal{L} = J_\alpha \times A_\alpha +  J_\beta \times A_\beta + \xi \parent{ \sum_k \lambda_k + \sum_l \lambda_l + \lambda_r+ \lambda_\alpha + \lambda_\beta - 1 } \, .
\label{L2}
\eea
The currents $J_\alpha= \lambda_\alpha/ \ell_\alpha$ and $J_\beta = \lambda_\beta/ \ell_\beta$ are fixed (as a reminder, $\ell_\alpha = \ell_\beta = 4$ denotes the length of the cycles).
Taking the derivates with respect to $\lambda_k, \lambda_l, \lambda_r$ and $\xi$, equating them to zero, and solving the resulting system yields \cite{FN03}
\bea
\lambda_k &=& J_\alpha (\rho-1) \quad \forall k\, , \label{Da} \\
\lambda_l &=&  J_\beta (\rho-1) \quad \forall l \, , \label{Db}\\
\lambda_r &=& \parent{ J_\alpha+ J_\beta } (\rho-1) \, , \label{Dc}
\eea 
where
\bea
\rho -1 = \frac{1-\ell_\alpha J_\alpha-\ell_\beta J_\beta}{\ell_\alpha J_\alpha+\ell_\beta J_\beta} \, .
\eea
We then obtain the minimal entropy production by inserting these quantities into the affinities $A_\alpha$ and $A_\beta$:
\bea
\Delta_i S_{\min} = \parent{\ell_\alpha J_\alpha+ \ell_\beta J_\beta} \ln \parent{1 + \frac{2(\ell_\alpha J_\alpha+\ell_\beta J_\beta)}{1-\ell_\alpha J_\alpha-\ell_\beta J_\beta} } \, .
\eea
By construction, this minimum is achieved by the dynamics (\ref{Da})-(\ref{Dc}).\\


\paragraph{Both currents have the same (clockwise) ortientations.}

In this case the affinities take the form
\bea
A_\alpha &=& \ln \parent{\frac{\lambda_r /2 + J_\alpha}{\lambda_r /2 + J_\beta}  } + \sum_k \ln \parent{1+\frac{2 J_\alpha}{\lambda_k}  } \, , \quad  k = [5,6], [6,1], [1,2]\\
A_\beta &=& \ln \parent{\frac{\lambda_r /2 + J_\beta}{\lambda_r /2 + J_\alpha} } + \sum_l \ln \parent{1+\frac{2 J_\beta}{\lambda_l}  } \, , \quad  l = [2,3], [3,4], [4,5]
\eea 
Using these affinities in the Lagrangian (\ref{L2}) and finding the corresponding minimum, we obtain
\bea
\lambda_k &=& J_\alpha (\rho-1) \quad \forall k\, , \label{Ca} \\
\lambda_l &=&  J_\beta (\rho-1) \quad \forall l \, ,\\
\lambda_r &=&  \max [ 0, \rho \, \vert J_\alpha - J_\beta \vert -\parent{ J_\alpha + J_\beta } ] \, , \label{Cc}
\eea
where $\rho$ is such that $\sum_k \lambda_k + \sum_l \lambda_l +\lambda_r = 1-\lambda_\alpha -\lambda_\beta$. 
This leads to
\bea
1/\rho =
\begin{cases}
 J_\alpha (\ell_\alpha-1) + J_\beta (\ell_\beta-1) + \vert J_\alpha - J_\beta \vert \quad \quad \quad \quad {\rm if} \quad \lambda_r > 0\\
[ J_\alpha (\ell_\alpha-1) + J_\beta (\ell_\beta-1) ] / (1-J_\alpha-J_\beta) \quad \quad {\rm otherwise} \, .
\end{cases}
\eea
The case $\lambda_r < 0$ occurs when both currents $J_\alpha$ and $J_\beta$ have similar values, $|J_\alpha - J_\beta| < K$ for some $K$.
  
\begin{figure}
\begin{center}
\includegraphics[scale=.55]{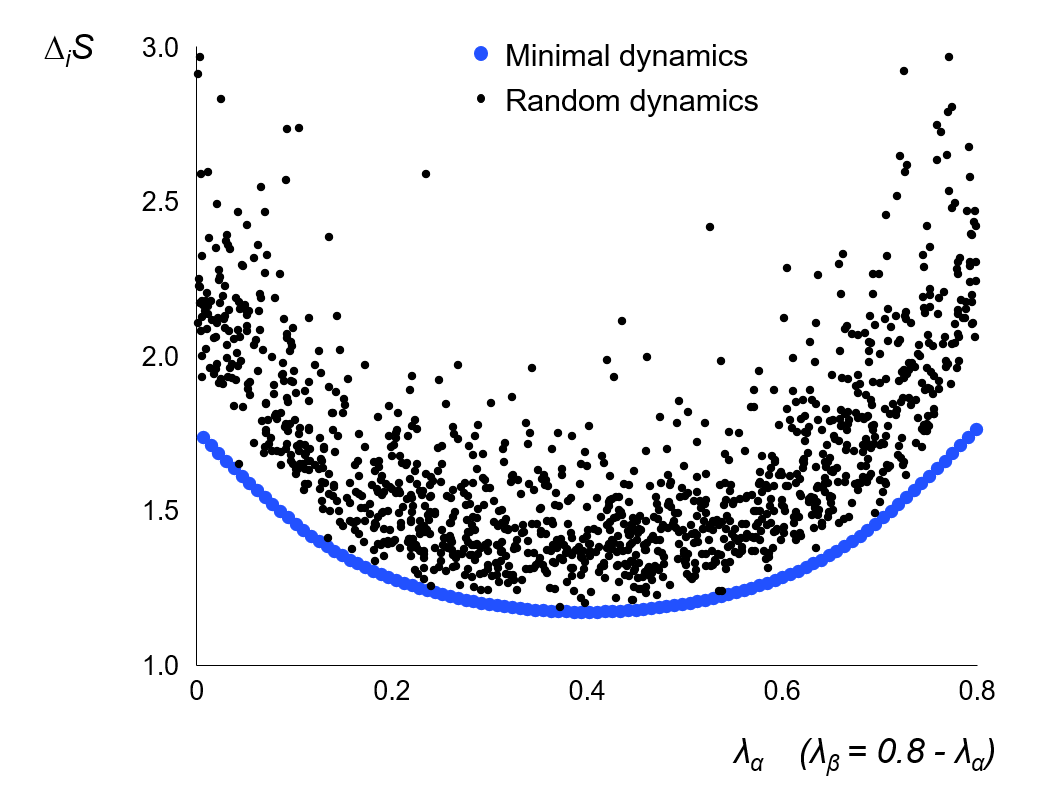}
\caption{{\bf Minimal entropy production for the transition network of Fig. (\ref{Ion})} when both currents flow in the same direction.
The minimal entropy production is shown for different values of the currents (blue).
Each black dot represents the entropy production of a random matrix $F$, numerically confirming our result (\ref{Ca}-\ref{Cc}).
We parametrize the currents with $\lambda_\alpha = J_\alpha \ell_\alpha$ and $\lambda_\beta = J_\beta \ell_\beta$, and take $\lambda_\beta = 0.8-\lambda_\alpha$.}
\label{EP2}
\end{center}
\end{figure}

We then obtain the minimal entropy production by inserting these quantities into the affinities $A_\alpha$ and $A_\beta$.
We numerically confirm the minimal entropy production as a function of the currents by comparing it with random Markov chains (Figure \ref{EP2}).

\subsection{General transition network}

We now consider the case of a Markov chain with an arbitrary transition network. 
We denote each edge in the transition network by $e$ and assign an orientation to each edge. 
For a graph with $N$ vertices and $E$ edges, there exists $M = E-N+1$ independent thermodynamic currents $J_\alpha$.
We can then express the local fluxes $F_e^{\pm}$ as a linear combination of the cycle flux $\lambda_e$ and the independent currents $J_\alpha$ \cite{A83, K06, S76}:
\bea
F_e^{\pm} = \frac{\lambda_e}{2} + J_e^{\pm} = \frac{\lambda_e}{2} + \sum_\alpha \epsilon_e^{\pm} (C^\alpha) J_\alpha
\eea
where $\epsilon_e^{\pm}$ takes the value $1$ if the cycle $C^\alpha$ contains the edge $e$ in the positive ($\epsilon^+$) or negative ($\epsilon^-$) direction, and $0$ otherwise.
The current along edge $e$ is given by $J_e=F_e^{+}-F_e^{-}=J_e^{+}-J_e^{-}$. 

We can then write the entropy production as
\bea
\Delta_i S = \sum_e  \parent{ F_e^{+} - F_e^{-} } \ln \parent{ \frac{F_e^{+}}{F_e^{-}}  }  = \sum_e  \parent{ J_e^{+} - J_e^{-} } \ln \parent{ \frac{ \lambda_e/2 + J_e^{+}}{\lambda_e/2 + J_e^{-}}  } \, ,
\label{disgeneral}
\eea
where $J_e^{\pm}$ are linear combinations of the (fixed) thermodynamic currents. 
Following the previous sections, we solve the associated Lagragian to obtain the minimal entropy production \cite{FN06}. 
The minimum is given by the dynamics
\bea
\lambda_e =\max \Big[ 0,  \rho \, \vert  J_e^{+} - J_e^{-} \vert -\parent{ J_e^{+} + J_e^{-} }  \Big]
\label{le}
\eea
where the normalization factor $\rho$ is obtained by solving the linear equation $\sum_e \lambda_e = 1-\sum_\alpha \lambda_\alpha$.

Inserting (\ref{le}) into (\ref{disgeneral}) we obtain the minimal entropy production of an arbitrary network given the thermodynamic currents $J_\alpha$.


This result takes into account the topology of the transition network and the thermodynamic currents. 
In other words, we derived the minimal entropy production required to sustain a given combination of thermodynamic currents in a given network topology. 
This result can also be used to study how different network topologies impact the entropy production.





\begin{thebibliography}{99}
\bibitem{FN02} This is a minimum as the affinity is not upper bounded for a given current: Taking any $\lambda_k = 0$ results in an irreversible chain with finite current but infinite affinity.
\bibitem{FN03} The derivation is left as an exercise. For reference, the Lagrange multiplier takes the value $\sqrt{1+2/\xi} = \rho-1$.
\bibitem{FN06} We consider networks with no self-transitions, $P_{ii} = F_{ii} =0$. This is not restrictive since minimizing the entropy production yields an optimum solution with $\lambda_{[i]} = 0$.
\end{thebibliography}

\vskip 1 cm

{\bf Disclaimer.} This paper is not intended for journal publication.

\end{document}